\documentclass[12pt]{article}
\begin{document}

\begin{center}
{\bf On the dual variables description of Yang-Mills theory}\\
\vspace{0.5cm}
A.G. Shuvaev\\

Petersburg Nuclear Physics Institute, Gatchina, S.Petersburg,
188300,
Russia\\

\end{center}

\begin{abstract}
The partition function of four dimensional $SO(4)$ Yang-Mills theory
is rewritten in terms of variables admitting straightforward relation
to the partition function of pure 4D gravity. The gauge action turns into
first-order Hilbert-Palatini action for Einstein gravity with a simple
extra term added. The proposed relation can be substantiated as a duality
for the partition functions provided a special gauge is imposed for
the gravity. The same method allows to find a closed expression for
the partition function of the $SO(4)$ gauge theory.
\end{abstract}

\section{Introduction}

Among modern approaches developing re\-la\-ti\-on\-ship bet\-ween
gra\-vi\-ty and Yang-Mills theory there are those based on
the first order formulation of general relativity
\cite{Plebanski, Halpern, Krasnov}.
They reformulate the gravity in terms of spin connection rather than
spacetime metric \cite{Capovilla} and provides a framework to treat
it as diffeomorphism invariant gauge theory, identifying the spin
connection with $SU(2)$ gauge field.
On the other hand the first order formalism enables to rewrite
Yang-Mills theory through gauge-invariant dual variables
\cite{Lunev_1, Lunev_2, Ganor:1995em, Diakonov:2001xg},
which is of interest from the viewpoint of unified description
of all interactions. The gravity action is quite different compared
to those of Yang-Mills theories that are believed to be responsible
for strong, weak and electromagnetic forces.
Therefore if there exists a change of variables that brings
the action of the gauge theory to the form close to the Einstein-Hilbert
one or vice versa it could be a way to unify both the theories.
Besides, such a transformation is interesting in itself as getting
further insight into the nature of Yang-Mills theory.

The variable $B_{\mu\nu}$ dual to the gauge field $A_\mu$ is introduced
as an auxiliary variable in the action \cite{Halpern, Diakonov:2001xg}
\begin{equation}
\label{SB}
S_B\,=\,\int d^4x \left(-\frac{g^2}{2}
{\rm Tr}\,B_{\mu\nu}B_{\mu\nu} + \frac{i}{2}
\varepsilon^{\mu\nu\lambda\sigma}
{\rm Tr}\,B_{\mu\nu}G_{\lambda\sigma}\right),
\end{equation}
which, being varied with respect $B_{\mu\nu}$, returns the Yang-Mills
action. Here $G_{\mu\nu}$ is the standard Yang-Mills field strength and
$\varepsilon^{\mu\nu\lambda\sigma}$ is antisymmetric tensor.
The basic feature of this form is that it makes the integral
over the field $A_\mu$ to be a Gaussian. It can be done explicitly
with the result given by the action
\begin{equation}
\label{BFaction}
S_{BF}\,=\,\frac{i}{2}\int d^4x
\varepsilon^{\mu\nu\lambda\sigma}
{\rm Tr}\,B_{\mu\nu}G_{\lambda\sigma}(\overline A)
\end{equation}
evaluated at the solution $\overline A_\mu$ of the equation of motion,
$\varepsilon^{\mu\nu\lambda\sigma}\bigl[D_\nu(\overline A),
B_{\lambda\sigma}\bigr]\,=\,0$, where $D_\nu(A)$ is a covariant
derivative. The action $S_{BF}$ turns out to be a functional of
the variable $B_{\mu\nu}$, which can be shown to be related
to the metric of the dual color space $g_{\mu\nu}$ and some
special extra metric $h_{ij}$ associated to $SU(2)$ gauge group.
Taken alone $S_{BF}$ defines topological "BF theory" \cite{BF}
If, in addition, the metric $h_{ij}$ is assumed
to be trivial, $h_{ij}=\delta_{ij}$, the action (\ref{BFaction})
reduces to the standard Einstein-Hilbert one $R\sqrt g$.
The first term in the total action breaks the general invariance
and plays the role of 'aether' \cite{Diakonov:2001xg}.

The starting point of this paper is to chose as dual variables
the set of fourvectors $e_\mu^A$ including $4\times 4 =16$
independent components. Upon putting
$B_{\mu\nu}^{AB}=e_\mu^A e_\nu^B - e_\nu^A e_\mu^B$
the second term in the formula (\ref{SB}) turns into Hilbert-Palatini
action with $e_\mu^A$ playing role of tetrad. At first glance,
dealing with $e_\mu^A$ we loss at once the gauge invariance of
the dual variables as well as the Gauss integral over them.
Indeed, they are the vectors under the gauge transformation,
$e_\mu^A(x)\to R^{AB}(x)e_\mu^B(x)$, where $R\in SO(4)$ in Euclidean
case (it would be a local Lorentz transformation in Minkowski space).
Substituting tetrad in the term $B_{\mu\nu}^2$ makes the integral
over $e_\mu^A$ to be non-Gaussian. The important fact however is that
there are no derivatives of the $e_\mu^A$ fields in it, so it is just
the product of usual finite dimensional integrals at each point $x$.
It is the main property the paper is based on. It enables to calculate
the integral and to relate the result with the partition function of
the gauge field. Taking the same integrals in the opposite order and
starting at first with the integral over field $A_\mu$,
which is still Gaussian, we arrive at the expression, which turns into
gravity action, when the redundant gauge degrees of freedom inherited
in the vectors $e_\mu^A$ are integrated out. These topics are discussed
in the Sections 2 and 3. The Section 4 is devoted to the modification
of these results due to various forms of quantum measure adopted
for gravity functional integral. It is argued in the Section 5 that
the additional term appearing in the gravity action and spoiling
its general coordinate invariance can be naturally interpreted
as gauge fixing. It allows to establish gauge/gravity connection,
or duality, at least for the partition functions.

It is shown in the Section 6 that the partition function of $SO(4)$
gauge field can be rewritten through the dual variables in a way,
where it turns completely into the product of independent integrals
at each point $x$. Thus the gluon partition function looks like
that calculated for ensemble of uncorrelated objects.

Note lastly that $SO(4)$ partition function we have dealt with throughout
the paper is simply related to $SU(2)$ one, $Z_{SO(4)}=Z_{SU(2)}^{\,2}$.

\section{Integral over tetrad}
Here we outline the main idea leaving the details as well as some
important modifications for the next section. We take the $SO(4)$
gauge group with the gauge field $A_\mu=A_\mu^{AB}T^{AB}$, where
$\mu=1,\ldots,4$ and $T^{AB}$, $A,B=1,\ldots,4$ are the generators
of 4D rotations, $T^{AB}=-T^{BA}$. The field strength tensor reads
$$
G_{\mu\nu}^{AB}(A)\,=\,\biggl(\partial_\mu A_\nu
- \partial_\mu A_\nu + \bigl[A_\mu,A_\nu\bigr]\biggr)^{AB}.
$$
Introducing auxiliary fields $e_\mu^A(x)$ playing further the role
of tetrad and tensor
$$
\Sigma_{\mu\nu}^{AB}\,=\,e_\mu^A e_\nu^B\,-\,
e_\nu^A e_\mu^B
$$
we present the partition function for the $SO(4)$ gauge field
through the following functional integral
\begin{equation}
\label{Ze}
Z\,=\,\int DA_\mu De_\mu^A\,\exp\int d^4x
\left[\,-\mu^4(e_\mu^A e_\mu^A)^2\,
+\,iM^2\,\widetilde G(A)\cdot \widetilde \Sigma\,\right],
\end{equation}
where
$$
\widetilde G(A)\cdot \widetilde \Sigma\,\equiv\,
\varepsilon^{\mu\nu\lambda\sigma}\varepsilon^{ABCD}
G_{\mu\nu}^{AB}(A)\,\Sigma_{\lambda\sigma}^{CD},
$$
and $\mu$ and $M$ are two arbitrary mass parameters.
To prove that the expression (\ref{Ze}) is really coincides with
the gluon partition function we directly calculate the functional
integral over tetrad. For this purpose we shall treat
it as a limit of multiple integral over discretized space,
\begin{eqnarray}
\label{ZAdis}
Z[A]\,&=&\,\prod_{x}\int de_\mu^A(x)\exp
\sum_{x}\left[-\mu^4\bigl(e_\mu^A(x) e_\mu^A(x)\bigr)^2\,
\Delta x^4 \right. \\
&&+\left.\,iM^2\widetilde G(x)\cdot \widetilde \Sigma(x)\Delta x^4
\right] \nonumber
\end{eqnarray}
($G(x)=G(A(x))$, $\Sigma(x)=\Sigma(e_\mu^A(x))$).
When the separation $\Delta x \to 0$, the multiplicity,
that is the number of finite-dimensional integrals located
at points $x$, goes to infinity while the Riemann sum turns
into continuous integral for the action in the exponent.

The crucial property the subsequent analysis is based on
is the absence of derivatives of the auxiliary fields $e_\mu^A(x)$
in the action. It makes the integrations over $e_\mu^A(x)$
to be independent from each other. In this context
the integral (\ref{ZAdis}) can be naturally thought of as
an averaging of the action functional over ensemble
of uncorrelated random variables,
$$
Z[A]\,=\,\langle\langle \exp\sum_{x}
iM^2\,\widetilde G(x)\cdot \widetilde \Sigma(x)\,\Delta x^4
\rangle\rangle,
$$
with the function
$$
P(e_\mu^A)\,=\,\exp\bigl[-\mu^4\,(e_\mu^A e_\mu^A)^2\bigr]
$$
providing distribution of these variables at each point.

After rescaling $e_\mu^A \to e_\mu^A/\mu\Delta x$ we get
\begin{eqnarray}
\label{Zprod}
Z[A]\,&=&\,C_0\,\prod_{x}\int de_\mu^A(x)\exp
\sum_{x}\left[-\bigl(e_\mu^A(x)\,e_\mu^A(x)\bigr)^2 \right. \\
&&+\,\left. i\frac{M^2}{\mu^2}\widetilde G(x)\cdot \widetilde \Sigma(x)
\Delta x^2\right], \nonumber
\end{eqnarray}
where the constant factor $C_0=\prod_x(\mu\Delta x)^{-16}$ is determined
by the number of lattice cells, $N=V_4/\Delta x^4$,
in the total space volume $V_4$. Introducing the notation
for independent averaging over tetrad at separate point,
\begin{equation}
\label{We}
\langle F \rangle \,=\,\int de_\mu^a e^{-(e_\mu^A e_\mu^A)^2}
F(e_\mu^A)
\end{equation}
the product takes the form
\begin{eqnarray}
\label{ZA}
Z[A]\,&=&\,C_0\,\prod_{x}\,\bigl[\,\langle \,1\, \rangle\, +\,
i\frac{M^2}{\mu^2}\langle\, \widetilde G(x)\cdot
\widetilde \Sigma(x) \,\rangle\Delta x^2 \\
&&- \frac 12\frac{M^4}{\mu^4}\langle \,(\widetilde G(x)
\cdot \widetilde \Sigma(x))^2
\rangle \Delta x^4 + {\cal O}\,(\Delta x^4)\,\bigr].
\nonumber
\end{eqnarray}
An apparent $O(16)$ symmetry of the weight integral in (\ref{We})
entails simple angular averaging. Combining the index pair
into a single multiple index $\alpha = \left\{A \atop \mu\right\}$
we have for $D=16$
\begin{eqnarray}
\label{ee}
\langle e_\alpha e_\beta\rangle\,&=&\,\langle e^2 \rangle
\delta_{\alpha\beta}\frac 1D  \\
\langle e_\alpha e_\beta e_\gamma e_\delta\rangle\,
\label{eeee}
&=&\,\langle e^4 \rangle
\bigl(\delta_{\alpha\beta}\delta_{\gamma\delta}
+\delta_{\alpha\gamma}\delta_{\beta\delta}
+\delta_{\alpha\delta}\delta_{\beta\gamma}\bigr)
\frac 1{D(D+2)},
\end{eqnarray}
$$
e^2\,\equiv\,e_\mu^A e_\mu^A,\quad
e^4\,\equiv\,(e_\mu^A e_\mu^A)^2.
$$
As a consequence, the second term in (\ref{ZA})
vanishes because of antisymmetry with respect to either
space or color indices. The third term yields
$$
-\frac 29\frac{M^4}{\mu^4}\,\langle e^4 \rangle \,
G_{\mu\nu}^{AB}(x)G_{\mu\nu}^{AB}(x)\,
\Delta x^4,
$$
so that
\begin{eqnarray}
Z[A]\,&=&\,Z_0\,\prod_{x}\bigl[1 - \frac 29\frac{M^4}{\mu^4}
\frac{\langle e^4 \rangle}{\langle\, 1\, \rangle}G^2(x)
\Delta x^4 + {\cal O}(\Delta x^4)\bigr] \nonumber \\
&=&\,Z_0\,\exp\left[- \frac 29\frac{M^4}{\mu^4}
\frac{\langle e^4 \rangle}{\langle\, 1\, \rangle}
\sum_{x}G^2(x)\Delta x^4 + {\cal O}(\Delta x^4)\right],
\nonumber
\end{eqnarray}
where $G^2 \equiv G_{\mu\nu}^{AB}G_{\mu\nu}^{AB}$ and normalization
factor $Z_0=\prod_x[(\mu\Delta x)^{-16}\cdot\langle\, 1\, \rangle]$.
Since the last line in the above expression is the integral sum,
we finally obtain for $\Delta x \to 0$
\begin{equation}
\label{ZAfin}
Z[A]\,=\,Z_0\,\exp\left[- \frac 29\frac{M^4}{\mu^4}
\frac{\langle e^4 \rangle}{\langle\, 1\, \rangle}
\int d^4x\,G^2(x)\right].
\end{equation}
Substituting here $\langle e^4 \rangle/ \langle 1 \rangle
=\int_0^\infty dr r^{19}e^{-r^4}/\int_0^\infty dr r^{15}e^{-r^4}=4$,
$r^2=e_\mu^Ae_\mu^A$, we arrive at the desired relation of
the functional integral (\ref{Ze}) to partition function of the $SO(4)$
gauge field,
$$
Z\,=\,Z_0\,\int DA_\mu\,\exp\left[-\frac 1{g^2}\int d^{\,4}x\,
G_{\mu\nu}^{AB}(A)\,G_{\mu\nu}^{AB}(A)\right],
$$
the coupling constant being
\begin{equation}
\label{g2}
\frac 1{g^2}\,=\,\frac 89\frac{M^4}{\mu^4}.
\end{equation}

\section{Relation to gravity}
There is a different way to work out the integral (\ref{Ze})
starting from the Gaussian integral over gluon fields.
We begin by noting that
$$
\varepsilon^{\mu\nu\lambda\sigma}\varepsilon^{ABCD}
\Sigma_{\lambda\sigma}^{\,CD}
=\,4\,\det(e)\cdot \bigl(e^{A,\mu}e^{B,\nu}-e^{B,\mu}e^{A,\nu}\bigr)\,
\equiv\,4\det(e)\,\Sigma^{\,AB,\mu\nu},
$$
where contravariant tetrad and metric tensor $g_{\mu\nu}$
are defined according to the relations
\begin{equation}
\label{eg}
g_{\mu\nu}\,=\,e_\mu^A e_\nu^A,~~~~
e_\mu^A\,=\,g_{\mu\nu}e^{A,\nu},~~~~
e^{A,\mu}\,e_\mu^B\,=\,\delta^{AB},~~~~
\det g\,=\,\det(e)^2.
\end{equation}
With these notations the second term in the exponent (\ref{Ze})
reads
\begin{equation}
\label{SeA}
iM^2\int d^4x\,\widetilde G(A)\cdot \widetilde \Sigma\,=\,
4iM^2\,\int d^4x\,\det(e)\,
G_{\mu\nu}^{AB}(A)\Sigma^{\,AB,\mu\nu},
\end{equation}
The expression (\ref{SeA}) is well-known Hilbert-Palatini action
(in Euclidean space), whose variation with respect $A_\mu$ and
$e^{A,\,\mu}$ yields General Relativity classical equations for pure
gravity \cite{Peldan}.

It is instructive here to pursue this connection in a little bit different
manner more suitable to carry out Gaussian integral.
To this end we first introduce covariant derivative, which acts
onto tetrad as
$$
\nabla_\mu e^{A,\nu}\,=\,\omega_\mu^{CA}e^{C,\nu},
$$
with the spin connection matrix $\omega_\mu^{AB}=-\omega_\mu^{BA}$.
It automatically implies metric compatibility, $\nabla_\lambda g_{\mu\nu}=0$,
and allows for the obvious identity
$$
\partial_\mu\bigl[\,\det(e)\Sigma^{\,AB,\mu\nu}A_\nu^{AB}\bigr]
-\partial_\nu\bigl[\,\det(e)\Sigma^{\,AB,\mu\nu}A_\mu^{AB}\bigr]
$$
$$
\,=\,\det(e)\nabla_\mu\bigl(\Sigma^{\,AB,\mu\nu}A_\nu^{AB}\bigr)
-\det(e)\nabla_\nu\bigl(\Sigma^{\,AB,\mu\nu}A_\mu^{AB}\bigr)
$$
$$
=\,\det(e)\bigl[A_\nu^{AB}\nabla_\mu\Sigma^{\,AB,\mu\nu}
-A_\mu^{AB}\nabla_\nu\Sigma^{\,AB,\mu\nu}
+\Sigma^{\,AB,\mu\nu}(\partial_\mu A_\nu-\partial_\nu A_\mu)^{AB}\bigr].
$$
Furthermore, we have
$$
A_\nu^{AB}\nabla_\mu\Sigma^{\,AB,\mu\nu}
-A_\mu^{AB}\nabla_\nu\Sigma^{\,AB,\mu\nu}\,=\,
2\,\bigl[\,\omega_\mu,\,A_\nu\bigr]^{AB}\Sigma^{\,AB,\mu\nu}.
$$
These two identities permit the field strength tensor
to be recast in the form
\begin{eqnarray}
\det(e)\Sigma^{\,AB,\mu\nu}G_{\mu\nu}^{AB}(A)&\,=\,&
\partial_\mu\bigl[\,\det(e)\Sigma^{\,AB,\mu\nu}A_\nu^{AB}\bigr]
-\partial_\nu\bigl[\,\det(e)\Sigma^{\,AB,\mu\nu}A_\mu^{AB}\bigr]
\nonumber \\
&\,+\,&\det(e)\,\Sigma^{\,AB,\mu\nu}
\biggl(\bigl[A_\mu-\omega_\mu,\,A_\nu-\omega_\nu\bigr]
-[\,\omega_\mu,\,\omega_\nu\bigr]\biggr)^{AB}\nonumber
\end{eqnarray}
valid for an arbitrary field $A_\mu$. Combining it with the same
expression written for $G_{\mu\nu}(\omega)$ we reach a net result:
\begin{eqnarray}
\label{Aomega}
&&\det(e)\,\Sigma^{\,AB,\mu\nu}G_{\mu\nu}^{AB}(A) \\
&&\,=\,
\partial_\mu\biggl[\det(e)\Sigma^{\,AB,\mu\nu}\bigl(A_\nu-\omega_\nu\bigr)^{AB}\biggr]
-\partial_\nu\biggl[\det(e)\Sigma^{\,AB,\mu\nu}\bigl(A_\mu-\omega_\mu\bigr)^{AB}\biggr]
\nonumber \\
&&\,+\,\det(e)\,\Sigma^{\,AB,\mu\nu}
\biggl(\bigl[A_\mu-\omega_\mu,\,A_\nu-\omega_\nu\biggr]
+G_{\mu\nu}(\omega)\biggr)^{AB}. \nonumber
\end{eqnarray}

It is worth sometimes to split second rank tensors into their self- and
anti-self-dual parts with respect to flat indices,
$$
T^{AB}\,=\,\stackrel{+~~~}{T^{AB}}\,+\,\stackrel{-~~~}{T^{AB}},~~~~
\stackrel{\pm~~~}{T^{AB}}\,=\,\frac 12\bigl(T^{AB}\pm
\varepsilon^{ABCD}T^{\,CD}\bigr),
$$
$$
\bigl[\,\stackrel{+}{T}_1,\,\stackrel{-}{T}_2\bigr]\,=\,0,~~~~
\stackrel{+~~~}{T_1^{AB}}\,\stackrel{-~~~}{T_2^{AB}}\,=\,0,
$$
the identity $[\tilde T_1,\tilde T_2]=[T_1,T_2]$,
$\tilde T^{AB}\equiv\varepsilon^{ABCD}T^{CD}$ being responsible for commutator
vanishing in the second line. In fact, this amounts to decomposition of
$SO(4)$ algebra into two $SU(2)$ algebras whose generators are made of
plus or minus components.

Substituting $A=A^\pm$ into equality (\ref{Aomega}) we immediately get
that it holds separately for plus and minus parts of the field strength
tensor,
\begin{eqnarray}
\label{Aomegapm}
&&\det(e)\,\Sigma^{\,AB,\mu\nu}G_{\mu\nu}^{AB}(A^\pm) \\
&&\,=\,
\partial_\mu\biggl[\det(e)\Sigma^{\,AB,\mu\nu}
\bigl(A_\nu^\pm-\omega_\nu^\pm\bigr)^{AB}\biggr]
-\partial_\nu\biggl[\det(e)\Sigma^{\,AB,\mu\nu}
\bigl(A_\mu^\pm-\omega_\mu^\pm\bigr)^{AB}\bigr]
\nonumber \\
&&\,+\,\det(e)\,\Sigma^{\,AB,\mu\nu}
\biggl(\bigl[A_\mu^\pm-\omega_\mu^\pm,\,A_\nu^\pm-\omega_\nu^\pm\bigr]
+G_{\mu\nu}(\omega^\pm)\biggr)^{AB} \nonumber
\end{eqnarray}
provided we take into account that
$G_{\mu\nu}^{AB}(A^\pm)\,=\,\stackrel{\pm~~~}{G_{\mu\nu}^{AB}}(A)$.

By virtue of the identity (\ref{Aomega}) the Gaussian integral over $A$
in (\ref{Ze}) can be trivially done by replacement $(A-\omega) \to \bar A$.
Moreover since the quadratic in fields $\bar A_\mu$ part of the action (\ref{SeA}),
\begin{eqnarray}
\label{K}
i\bar A\cdot {\cal K}\cdot \bar A\,&=&\,4iM^2\int d^4x\,\det(e)\,
\Sigma^{\,AB,\mu\nu}\,
\bigl[\bar A_\mu,\,\bar A_\nu\bigr]^{AB} \\
\,&=&\,4iM^2\int d^4x\,\det(e)\,
\Sigma^{\,AB,\mu\nu}\,
\bigl(\bigl[\bar A_\mu^+,\,\bar A_\nu^+\bigr]
+\bigl[\bar A_\mu^-,\,\bar A_\nu^-\bigr]\bigr)^{AB} \nonumber
\end{eqnarray}
does not contain derivatives, the functional determinant decays
into infinite product of usual determinants at each space point,
\begin{equation}
\label{detK}
{\rm Det}({\cal K}/\pi)^{-\frac 12}\,=\,\prod_{x} c\,\det[e(x)]^{-6}
\bigl[M^2 \Delta x^4\bigr]^{-12}
\end{equation}
($c=2^{-28}\pi^{12}$), producing
an additional local factor for the functional measure $De_\mu^A$.
Thus the integral (\ref{Ze}) takes the form
\begin{equation}
\label{Z}
Z\,=\,\int De_\mu^A\,\rho(e_\mu^A)\exp\int d^4x
\left[\,-\mu^4(e_\mu^A e_\mu^A)^2\,
+\,4iM^2\,G_{\mu\nu}^{AB}\Sigma^{\,AB,\mu\nu}\right].
\end{equation}
Recalling that
$$
4iM^2\!\int d^4x \,G_{\mu\nu}^{AB}\Sigma^{\,AB,\mu\nu}\,=\,
8iM^2\int d^{\,4}x\,R\det{e}
$$
looks like the conventional Einstein-Hilbert action, this form bears
a close resemblance to gravity partition function. A straightforward
identification with it requires, however, a physical interpretation
of the first term, which explicitly spoils general covariance of
the action appearing in the integral (\ref{Z}). Further, the space
volume in the Einstein-Hilbert action should be positive,
$\sqrt{g}=|\det(e)|$, which has to impose certain restrictions on
the integrals over tetrad. Another possible
question is a proper choice of the local factor $\rho(e_\mu^A)$.
These issues will be discussed in the next two sections.

\section{Gravity measure}
For proper description of quantum gravity the functional
(\ref{Ze}) needs somewhat modification, namely, the functional
measure $De_\mu^A$ has to be changed
by extra local factor.

There are several alternative ways to define its form.

Similarly to gauge fields theories, where
the measure $DA_\mu$ is invariant under gauge transformation,
the relevant gravity measure $Dg$ is supposed to be invariant under
general coordinate transformation. The invariance is achieved
by inclusion of a local factor \cite{KP},
\begin{equation}
\label{Dg}
Dg\,=\,\prod_x\prod_{\mu\le\nu}g^{\frac 52}\,dg^{\,\mu\nu}\,=\,
\prod_x\prod_{\mu\le\nu}g^{-\frac 52}\,dg_{\mu\nu},
\end{equation}
$g=\det(g_{\mu\nu})$.

Another form has been obtained in the hamiltonian formalism,
which ensures unitarity of $S$-matrix for gravity \cite{FradVilk},
\begin{equation}
\label{DgFV}
Dg\,=\,\prod_x\prod_{\mu\le\nu}g^{-\frac 32}\,g^{00}dg_{\mu\nu}
\end{equation}

The hamiltonian formalism applied to the Palatini-Holst action
leads to the measure \cite{Engle}, which in our notations reads
\begin{equation}
\label{VVs}
DA\,De\,=\,\prod_x dA_\mu^{AB} de_\mu^A\,{\cal V}^3V_s,
\end{equation}
where ${\cal V}=\sqrt{g}$ is a spacetime volume element while
$V_s$ is a spatial volume element.

Actually the local factor in the measures affects only
high orders of the perturbation theory acting as counterterm
that cancels the divergent pieces of the loops proportional to
$\delta^{(4)}(0)$\cite{FradVilk}. The measures (\ref{DgFV},
\ref{VVs}) are not invariant under coordinate transformations.
The reason for this lies in the procedure of quantization, namely,
in a spacetime lattice that is implied behind path integral.
It provides an ultraviolet regularization but violates invariance
to the coordinate transformation \cite{Leutwyler}.

Despite the modifications the steps (\ref{ZAdis}),
(\ref{Zprod}), (\ref{ZA}) hold unchanged provided
the averaging is redefined according to the measures
(\ref{Dg}),(\ref{DgFV}), (\ref{VVs}). Particular form
of the local factors results only in different coefficients
in the equation for the coupling constant (\ref{g2}).
To show it we begin with the measure
\begin{equation}
\label{DeK}
DA\,De\,=\,\prod_{x}dA_\mu^{AB}de_\mu^A\,\det(e)^K
\end{equation}

The local factor in (\ref{DeK}) requires to redefine
the averaging over tetrad as
\begin{equation}
\label{detWe}
\langle F \rangle_K\,=\,\int de_\mu^A \,\det(e)^K\,
e^{-(e_\mu^A e_\mu^A)^2}F(e_\mu^A).
\end{equation}
The additional factor $\det(e)^K$ breaks $O(16)$ symmetry and
invalidates the identities (\ref{ee}),(\ref{eeee}). But it allows
for independent $O(4)$ rotations of color and space tetrad indices,
that results into the set of relations instead of
(\ref{ee}),(\ref{eeee}):
\begin{equation}
\label{detee}
\langle e_\mu^A e_\nu^B\rangle_K\,=\,\delta^{AB}\delta_{\mu\nu}\,
\frac 1{16}\,\langle g_{\alpha\alpha}\rangle_K,
\end{equation}
\begin{equation}
\label{deteeee}
\langle e_\mu^A e_\nu^B e_\lambda^C e_\sigma^D\rangle_K\,=\,
\delta^{AB}\delta^{CD}s_{\mu\nu\lambda\sigma}
+ \delta^{AC}\delta^{BD}s_{\mu\lambda\nu\sigma}
+ \delta^{AD}\delta^{BC}s_{\mu\sigma\nu\lambda},
\end{equation}
with
\begin{equation}
\label{So}
s_{\mu\nu\lambda\sigma}=\frac 1{576}\,\bigl[(3g_1^2 - 2g_2)
\delta_{\lambda\sigma}\delta_{\mu\nu}
-(g_1^2 - 2g_2)(\delta_{\lambda\mu}\delta_{\nu\sigma}
+ \delta_{\lambda\nu}\delta_{\mu\sigma})\bigr]
\end{equation}
and
\begin{equation}
\label{g1g2}
g_1^2\,=\,\langle g_{\alpha\alpha}g_{\beta\beta}\rangle_K,~~~~~
g_2\,=\,\langle g_{\alpha\beta}g_{\beta\alpha}\rangle_K.
\end{equation}
The first identity (\ref{detee}) removes the $\Delta x^2$
term in the expansion (\ref{ZA}), while the second one
(\ref{deteeee}) returns the value of the term proportional
to $\Delta x^4$, which yields in the continuous limit
\begin{equation}
\label{detZAfin}
Z[A]\,=\,Z_0\,\exp\left[- \frac {4}{9}\frac{M^4}{\mu^4}
\frac{g_1^2-g_2}{\langle\, 1\, \rangle_K}
\int d^{\,4}x\, G^{\,2}(x)\right],
\end{equation}
$$
Z_0\,=\,\prod_x\bigl[(\mu\Delta x)^{-16-4K}\langle\,1\,\rangle_K\,
\bigr].
$$
Using $g_1^2-g_2$ value calculated in the Appendix
we obtain the coupling constant
\begin{equation}
\label{detg2}
\frac 1{g^2}\,=\,\frac {2}{3}\frac{M^4}{\mu^4}
\frac{(K+3)(K+4)}{2K+9}.
\end{equation}
and arrive at the relation
\begin{equation}
\label{Fine}
\int De\exp\int d^4x\left[-\mu^4 (e_\mu^A e_\mu^A)^2 +
8iM^2\,R\det(e)\,\right]
\end{equation}
$$
=\,Z_{gA}\,\int DA\,\exp\left[-\frac 1{g^2}\int d^{\,4}x\,
G_{\mu\nu}^{AB}(A)\,G_{\mu\nu}^{AB}(A)\right].
$$
The functional measure is understood as
$$
De\,=\,\prod_x\rho(e_\mu^A)de_\mu^A,
$$
with the local factor encountering the determinant
(\ref{detK}), $\rho(e_\mu^A)=\det(e)^{K-6}$.
The "transition" coefficient between gravity and gauge partition
function is given by the product running over space points,
\begin{equation}
\label{ZgA}
Z_{gA}\,=\,\prod_x c_e M^{-24}\left[\bigl(\mu^4\Delta x^4\bigr)^{-4-K}
\langle\,1\,\rangle_{K} \bigl(M^4\Delta x^4\bigr)^{12}
\right],
\end{equation}
$c_e=2^{28}\pi^{-12}$.

Now we have to correct this result for the condition
$\det(e)\ge 0$ assumed in the gravity action.
One way to incorporate it is to integrate over configurations
for which the inequality holds, that is to choose instead of
the measure (\ref{DeK}) the modified one,
\begin{equation}
\label{DeK+}
DA\,De^+\,=\,\prod_x dA_\mu^{AB}de_\mu^A\det(e)^K\,
\theta\bigl(\det(e)\bigr).
\end{equation}
Indeed, the functional, coming about upon the integration
over gauge field $A_\mu$, depends only on the metric tensor,
the curvature, that is expressed through the metric tensor again,
and $\det(e)$. The tensor $g_{\mu\nu}=e_\mu^A e_\nu^A$ remains
unchanged under reflection of sign in any row of the matrix $e_\mu^A$,
that negates $\det(e)$. Therefore for an arbitrary function of these
variables one may write
\begin{equation}
\label{theta}
\int de_\mu^A\,\theta\bigl(\det(e)\bigr) \,f(\det(e),g_{\mu\nu})\,=\,
\frac 12 \int de_\mu^A\,\,f(\sqrt{g},g_{\mu\nu}).
\end{equation}
It allows to pass from tetrad to the integral over $g_{\mu\nu}$
with the measure
\begin{equation}
\label{DgN}
Dg\,=\,\prod_x\prod_{\mu\le\nu}g^{N}\,dg_{\mu\nu}.
\end{equation}
The power $N$ is separated here into two parts. The first part
is due to the determinant (\ref{detK}), the second one arises
as a local Jacobian, since for any function $f$
$$
\int \prod_{\mu, A}de_\mu^A f(e_\mu^A e_\nu^A)\,=\,
\int\prod_{\mu\le\nu}\,dg_{\mu\nu} g^{-\frac 12}f(g_{\mu\nu})
$$
(see the Appendix). Putting it together with $\det(e)^K$
we recover the measure (\ref{DgN}) for $K=2N+7$.

The measure (\ref{DeK+}) admits $SO(4)\times SO(4)$ independent
rotations of color and space indices rather than $O(4)\times O(4)$ before.
This restriction allows for one more term in the equation (\ref{deteeee}),
\begin{eqnarray}
\langle e_\mu^A e_\nu^B e_\lambda^C e_\sigma^D\rangle_K\,&=&\,
\delta^{AB}\delta^{CD}s_{\mu\nu\lambda\sigma}
+ \delta^{AC}\delta^{BD}s_{\mu\lambda\nu\sigma}
+ \delta^{AD}\delta^{BC}s_{\mu\sigma\nu\lambda} \nonumber \\
\label{deteeee+}
&&\,+ \frac 1{24}\varepsilon^{ABCD}\varepsilon_{\mu\nu\lambda\sigma}
\langle\,\det(e)\,\rangle_K,
\end{eqnarray}
and the constraint $\det(e)\ge 0$ amounts to an extra contribution
to the gauge action,
\begin{eqnarray}
Z[A]\,&=&\,Z_0\,\exp\int d^4x\left[-\frac 1{g^2}
G_{\mu\nu}^{AB}(A)\,G_{\mu\nu}^{AB}(A)\right. \nonumber \\
&&+\,\left. \tilde c\,\varepsilon^{ABCD}\varepsilon_{\mu\nu\lambda\sigma}\,
G_{\mu\nu}^{AB}(A)\,G_{\lambda\sigma}^{CD}(A)\right],\nonumber
\end{eqnarray}
This term appears to be a total derivative
with the constant $\tilde c$ in front given by the averaged
'volume element'
$$
\tilde c\,=\,\frac 43\frac{\langle\,|\det(e)|\,\rangle_K}
{\langle\,1\,\rangle_K}\frac{M^4}{\mu^4},
$$
which for the particular form (\ref{DeK+}) evaluates to
$\tilde c=\frac 1{3\sqrt \pi}\frac{2K+1}{2K+9}$.

Consider now the measures (\ref{DgFV}) and (\ref{VVs}).
They share the common lack of explicit covariance
because of their hamiltonian nature that distinguishes
'time' and 'space'. It breaks the $O(4)$ symmetry
of the local averaging with respect to the space indices
though the $O(4)$ or $SO(4)$ color symmetry remains intact,
so that the relations (\ref{detee}) and (\ref{deteeee}) or
(\ref{deteeee+}) are still applicable while the formula
(\ref{So}) should be changed. Introducing 'time' directed
unit vector $n_\mu$ one can write instead of it
\begin{equation}
\label{Sn}
s_{\mu\nu\lambda\sigma}\,=\,
c_1\,\delta_{\mu\nu}\delta_{\lambda\sigma}\,
+\,c_2\,(\delta_{\lambda\mu}\delta_{\nu\sigma}
+ \delta_{\lambda\nu}\delta_{\mu\sigma})
\,+\,c_3\,n_\mu n_\nu n_\lambda n_\sigma,
\end{equation}
the coefficients being again expressed through averaged values
(\ref{g1g2}) and a new structure
$g_{nn}=\langle n_\alpha g_{\alpha\beta}n_\beta\rangle_K$,
\begin{eqnarray}
c_1\,&=&\,\frac 1{1512}(8g_1^2 - 5g_2 - 3g_{nn}),~~~~
c_2\,=\,-\frac 1{3024}(5g_1^2 - 11g_2 - 6g_{nn}),\nonumber \\
c_3\,&=&\,-\frac 1{504}(g_1^2 + 2g_2 - 24g_{nn}), \nonumber
\end{eqnarray}
where the averaging is carried out with particular local factors
determining the measure. The output gluon partition function
has the same form (\ref{detZAfin}), in which the details of
gravity measure are encoded only in the parameters $g_1^2$,
$g_2$, $\langle\,1\,\rangle$ and $\langle\,|\det(e)|\,\rangle$.
The last one defines the total derivative term appearing in
the gluon action, when the constraint $\det(e)>0$ is imposed.
The term $g_{nn}$ does not contribute.

Thus the variation of the functional measure affects
the final relation between gravity and gauge field,
\begin{equation}
\label{Fin}
\int Dg\exp\int d^4x\left[-\mu^4 g_{\mu\mu}^2 +
8iM^2\,R\sqrt{g}\,\right]
\end{equation}
$$
=Z_{gA}\,\int DA\,\exp\left[-\frac 1{g^2}\int d^{\,4}x\,
G_{\mu\nu}^{AB}(A)\,G_{\mu\nu}^{AB}(A)
+ \tilde c \int d^{\,4}x\,
G_{\mu\nu}^{AB}(A)\,\tilde G_{\mu\nu}^{AB}(A)
\right],
$$
only through coupling $g$, constant $\tilde c$ and normalization product (\ref{ZgA})
(with $c_e=2^{29}\pi^{-12}$ because of 1/2 in (\ref{theta})).
The average $\langle\,1\,\rangle_{K}$ is calculated for a given
local factor $\rho(g)$ in the gravity measure. It is supposed to be
homogenous function for the metrics rescaling, $g_{\mu\nu}\to \alpha g_{\mu\nu}$,
$\rho(g)\to \alpha^{4N}\rho(g)$, $K=2N+7$.

\section{Duality}
The second term in the l.h.s. (\ref{Fin}) is the conventional
Einstein-Hilbert action that enjoys general coordinate
invariance in contrast to the first term that explicitly
violates it. To clarify a meaning behind it we consider
partition function for the pure Euclidean gravity without
extra terms,
$$
Z_g\,=\,\int Dg\exp\int d^4x\left[\,8iM^2 R\,\sqrt{g}\,\right].
$$
This functional is to be supplemented with appropriate
gauge conditions fixing in gravity case the coordinate system.
The four coordinates should be fixed with four constraints
imposed on the metric tensor. Let us choose as variables subject
to the constraints the diagonal components of $g_{\mu\mu}$,
\begin{equation}
\label{gauge}
g_{\mu\mu}(x)\,=\,\alpha_\mu(x).
\end{equation}
According to the standard proceeding it amounts to dealing
with gauge-fixed integral
$$
Z_g^{gf}\,=\,\int Dg\,\Delta_{FP}[g]\prod_\mu
\delta\,[g_{\mu\mu}-\alpha_\mu]
\exp\int d^{\,4}x\left[\,8iM^2\,R\,\sqrt{g}\,\right]
$$
with Faddeev-Popov determinant $\Delta_{FP}[g]$.
Two integrals, $Z_g$ and $Z_g^{gf}$ differ only in the constant
normalization proportional to the volume of gauge group that is
the group of coordinate diffeomorphisms, in our case. The first
order variation of the gauge conditions under the infinitesimal
action of this group,
$$
\delta x_\mu \,=\, \epsilon_\mu(x), ~~~~
\delta g_{\mu\mu}(x)\,=\,-2\nabla_\mu \epsilon_\mu(x)
$$
yields Faddeev-Popov determinant,
$$
\Delta_{FP}[g]\,=\,\prod_\mu {\rm Det}\bigl(\nabla_\mu\bigr).
$$
Expressing covariant derivative through Christoffel symbols,
$(\nabla_\mu)_\beta^\alpha = \partial_\mu\delta_\beta^\alpha
+ (\Gamma_\mu)_\beta^\alpha$, we rewrite the determinant as
$$
{\rm Det}\,\partial_\mu\cdot {\rm Det}\,
\bigl[1+\theta\cdot\Gamma_\mu\bigr],
$$
where $\partial_\mu \theta(x_\mu-y_\mu)=\delta(x_\mu-y_\mu)$, or
$$
{\rm Det}\bigl(\nabla_\mu\bigr)\,=\,
{\rm Det}\,\partial_\mu\,\cdot\,\exp\bigl\{\,\sum_{n=1}^\infty
\frac{(-1)^{n+1}}{n}\,
{\rm Tr}\,\bigl[\theta\cdot\Gamma_\mu\,\bigr]^n\bigr\}.
$$
Apart the index summation the trace implies the loop integrals over
coordinates, which all are zero because of the arguments ordering
in $\theta(x-y)$ function. Thus
${\rm Det}\bigl(\nabla_\mu\bigr)={\rm Det}\,\partial_\mu$ is a constant
not depending on the metric tensor $g_{\mu\nu}$. This property
is similar to ghost decoupling well known for axial or planar gauges
in QCD \cite{Lb}. Furthermore, the functional $Z_g^{gf}$ does not
depend on the functions $\alpha_\mu$ since its variation
only "moves the point along the same gauge group orbit".
It allows to average $Z_g^{gf}$ over $\alpha_\mu$ with arbitrary
weight factor that results only in an overall constant in front,
\begin{equation}
\label{Phi}
Z_g^{gf}\,=\,N\int D\alpha_\mu\,\Phi[\alpha]
\int Dg\,\prod_\mu \delta\,[g_{\mu\mu}-\alpha_\mu]
\exp\int d^{\,4}x\left[\,8iM^2\,R\,\sqrt{g}\,\right]
\end{equation}
Choosing
$$
\Phi[\alpha]\,=\,\exp\bigl[-\mu^4\int d^4x (\sum_\nu\alpha_\nu)^2\bigr]
$$
we arrive at the functional (\ref{Fin}). One can conclude therefore
that there is a duality between the $SO(4)$ gauge theory and quantum
gravity taken in the particular gauge (\ref{gauge}).

\section{Further implications}
The weight function in the integral (\ref{Ze}) may be viewed,
by itself, as a part of local factor in the measure,
\begin{equation}
\label{Zrho}
Z\,=\,\int DA_\mu De_\mu^A\,\rho(e_\mu^A)\exp i\int d^4x
\,M^2\,\widetilde G(A)\cdot \widetilde \Sigma,
\end{equation}
$$
\rho(e_\mu^A)\,=\,\exp\{-(\mu\Delta x)^4\bigl(e_\mu^A e_\mu^A)^2\}.
$$
This form suggests a natural extension to other weight functions
that admit continuous limit, for example,
$$
\rho(e_\mu^A)\,=\,\exp\{-(\mu\Delta x)^4\bigl(e_\mu^A e_\mu^A)^2\}
+ c \exp\{-\alpha(\mu\Delta x)^4\bigl(e_\mu^A e_\mu^A)^2\}.
$$
The formulae (\ref{ZAfin}), (\ref{detZAfin}) still hold in this case
if the average values (\ref{We}), (\ref{detWe}) appearing in them
are modified in the same manner, that is $e^{-(e_\mu^A e_\mu^A)^2}
\to e^{-(e_\mu^A e_\mu^A)^2} + c e^{-\alpha (e_\mu^A e_\mu^A)^2}$.
Similarly, taking the functional $\Phi[\alpha]$ in (\ref{Phi})
as infinite product of local terms,
$$
\Phi[\alpha]\,=\,\prod_{x,\mu}\rho(e_\mu^A),
$$
we again reproduce gauge/gravity duality interpreting noncovariant
weight factor in the measure as an ingredient of gauge fixing procedure
for the gravitational field.

Another interesting perspective comes about if we replace
the seed functional integral $Z[A]$ (\ref{Ze}) by
the expression,
\begin{equation}
\label{Zeps}
Z_\varepsilon[A]\,=\,\int DA_\mu De_\mu^A\,\det(e)^K
e^{S_\varepsilon}
\end{equation}
with a new action,
$$
S_\varepsilon\,=\,\int d^4x
\left[\,-\mu^4(e_\mu^A e_\mu^A)^2\,
+\,iM^2\det(e)\,\varepsilon^{ABCD}\Sigma^{\,AB,\mu\nu}
G_{\mu\nu}^{CD}\,\right],
$$
and arbitrary power $K$ in the functional measure.
The action of the form
$$
S_H\,=\,\frac 14\int d^4x\,\det(e)\,
\Sigma^{\,AB,\mu\nu}\bigl[\,G_{\mu\nu}^{AB}
-\frac 1{2\gamma}\varepsilon^{ABCD}G_{\mu\nu}^{CD}\,\bigl]
$$
is generalized Hilbert-Palatini action proposed by Holst
\cite{Holst}. It gives rise to the same equation of motion
for classical gravity regardless the value of Immirzi
parameter $\gamma$ \cite{Immirzi} (though it may affect
quantum theory \cite{Rovelli}). It is this second,
Immirzi related term, in the Holst action that is only
left in the action $S_\varepsilon$ in (\ref{Zeps}).

Proceeding as before and integrating over tetrad
with the help of equalities
$$
\det(e)\,\varepsilon^{ABCD}\Sigma^{\,CD,\mu\nu}\,=\,
\varepsilon^{\mu\nu\lambda\sigma}\Sigma_{\lambda\sigma}^{AB}
$$
and (\ref{detee}), (\ref{deteeee})
we draw the connection of integral (\ref{Zeps})
to the partition function of $SO(4)$ gauge field
similar to (\ref{detZAfin}),
\begin{eqnarray}
\label{Zeps[A]}
Z_\varepsilon[A]\,&=&\,\prod_x\biggl[(\mu \Delta x)^{-16-4K}
\langle\,1\,\rangle_K\biggr]\,\\
&&\times\int DA\,\exp\biggl\{- \frac {1}{9}\frac{M^4}{\mu^4}
\frac{g_1^2-g_2}{\langle\, 1\, \rangle_K}
\int d^4x\,G^2\biggr\},\nonumber
\end{eqnarray}
which amounts to the coupling constant value
\begin{equation}
\label{gK}
\frac 1{g^2}\,=\,\frac{1}{6}\frac{(K+3)(K+4)}{2K+9}
\frac{M^4}{\mu^4}.
\end{equation}
On the other hand, presenting dual tensor as
$$
\det(e)\,\varepsilon^{ABCD}\Sigma^{\,AB,\mu\nu}
G_{\mu\nu}^{CD}\,=\,\det(e)
\bigl[\,\stackrel{+~~~}{G_{\mu\nu}^{AB}}\,-\,
\stackrel{-~~~}{G_{\mu\nu}^{AB}}\,\bigr]
\Sigma^{\,AB,\mu\nu},
$$
and recalling the identities (\ref{Aomegapm}),
we bring the action that appears in (\ref{Zeps}) to the form
(omitting total derivatives)
$$
S_\varepsilon\,=\,\int d^4x
\left[\,-\mu^4(e_\mu^A e_\mu^A)^2\,
+\,iM^2\,\det(e) \Sigma^{\,AB,\mu\nu}
\left(\bigl[A_\mu^+ -\omega_\mu^+,\,A_\nu^+ -\omega_\nu^+\bigr]
\right.\right.
$$
$$
- \left.\left. \bigl[A_\mu^- -\omega_\mu^-,\,
A_\nu^- -\omega_\nu^-\bigr]\right)^{AB}
\,+\,iM^2\,\det(e) \Sigma^{\,AB,\mu\nu}
\bigl[\,\stackrel{+}{G}_{\mu\nu}(\omega)\,
-\,\stackrel{-}{G}_{\mu\nu}(\omega)\,\bigr]^{AB} \right].
$$
The last term is identically zero here, as immediately follows
from the Bianchi identity for curvature tensor,
$$
\det(e)\Sigma^{\,AB,\mu\nu}\varepsilon^{ABCD}{G}_{\mu\nu}^{CD}(\omega)
\,=\,\frac 12\det(e)\Sigma^{\,AB,\mu\nu}R_{\lambda \sigma \mu \nu}
\Sigma^{\,CD,\lambda \sigma}
$$
$$
\,=\,\frac 12\varepsilon^{\mu\nu\alpha\beta}\Sigma_{\alpha\beta}^{CD}
R_{\lambda \sigma \mu \nu}\Sigma^{\,CD,\lambda \sigma}\,=\,
2\varepsilon^{\mu\nu\lambda\sigma}R_{\lambda \sigma \mu \nu}\,=\,0.
$$
As a consequence the replacement $A_\mu-\omega_\mu \to A_\mu$ completely
removes all derivatives from $S_\varepsilon$, so that the whole integral
(\ref{Zeps}) turns into product of independent integrals uncorrelated
at each space point both for tetrad and gauge fields,
\begin{eqnarray}
\label{Zepsf}
Z_\varepsilon[A]\,&=&\,\int DA_\mu De_\mu^A\,\det(e)^K
\exp \int d^4x\,\left[\,-\mu^4(e_\mu^A e_\mu^A)^2\right. \\
&&+\left. \,iM^2\,\det(e) \Sigma^{\,AB,\mu\nu}
\left(\bigl[A_\mu^+,\,A_\nu^+ \bigr]
- \bigl[A_\mu^-,\,A_\nu^- \bigr]\right)^{AB}\right]. \nonumber
\end{eqnarray}

The form (\ref{Zepsf}) entails, in particular, a closed expression
of the same type for the gluon partition function $Z[A]$.
Indeed, comparing (\ref{Zepsf}) with (\ref{Zeps}) and (\ref{Zeps[A]}),
we present $Z[A]$ through uncorrelated product as
\begin{equation}
\label{Z[A]_1}
Z[A]\,=\,\prod_x \int dA\exp\biggl\{-\frac 1{g^2}
\bigl[A_\mu\,,\,A_\nu\bigr]^{AB}
\bigl[A_\mu\,,\,A_\nu\bigr]^{AB} (\Delta x)^4
\biggr\}.
\end{equation}
This simple result is plagued by non-decreasing of the integrand
along $A_\mu = A_\nu$ directions, which makes it divergent at each
space point $x$. However it is finite if the integrals in (\ref{Zepsf})
are taken in opposite order. Integrating out at first
the gauge fields yields (after rescaling
$e_\mu^A \to e_\mu^A/(\mu \Delta x)$)
$$
Z_\varepsilon\,=\,\prod_x \bigl(\mu\Delta x\bigr)^{-16-4K}
\left[\int de_\mu^A \,\det(e)^K e^{-(e_\mu^Ae_\mu^A)^2}
\bigl(\det\frac{\tilde {\cal K}}{\pi}\bigr)^{-\frac 12}\right],
$$
the matrix $\tilde {\cal K}$ defining quadratic form in the Gaussian
integral,
$$
iA\cdot \tilde {\cal K}\cdot  A\,=\,iM^2\,\int d^4x\,\det(e)
\Sigma^{\,AB,\mu\nu}
\left(\bigl[A_\mu^+,\,A_\nu^+ \bigr]
- \bigl[A_\mu^-,\,A_\nu^- \bigr]\right)^{AB}.
$$
This form differs from that appearing in (\ref{K})
only in the relative sign between the positive and negative
blocks and the overall coefficient in front. Since both matrices,
${\cal K}$ and $\tilde{\cal K}$, are block diagonal,
their determinants are proportional, and
$$
Z_\varepsilon[A]\,=\,\prod_x\left[(\mu \Delta x)^{-16-4K}
\left(\frac{M}{\mu}\Delta x\right)^{-24}
\langle\,1\,\rangle_{K-6}\,\frac{\pi^{12}}{16}\right].
$$
Comparing again this expression with (\ref{Zeps[A]})
we arrive at the "finite" result for the gluon partition
function,
\begin{equation}
\label{Z[A]_2}
Z[A]\,=\,\prod_x \left[\frac{\pi^{12}}{16}\left(\frac{M}{\mu}
\Delta x\right)^{-24}\frac{\langle\,1\,\rangle_{K-6}}{\langle\,1\,\rangle_{K}}
\right].
\end{equation}
Substituting here explicit expressions for $\langle\,1\,\rangle_{K}$
from the Appendix and the coupling constant (\ref{gK}) we get
\begin{equation}
Z[A]\,=\,\prod_x \left[g^{12} (\Delta x)^{-24} c_K\right]
\end{equation}
with $c_K\,=\,4\pi^{15}
\left[\frac{(K+3)(K+4)}{3(2K+9)}\right]^6
\frac{(2K+7)(2K+5)(4K^2-9)(4K^2-1)}{(K-1)(K-3)(K-5)}$.

Thus the divergency in (\ref{Z[A]_1}) may be regarded as
a consequence of the continuous limit that has not been
assumed in deriving (\ref{Z[A]_2}). Obviously, the divergency
has an ultraviolet origin, since it appears when $\Delta x \to 0$.
The equation (\ref{Z[A]_2}) is then natural to treat
as being obtained with a kind of lattice regularization
characterized, besides the fixed $\Delta x$, by the parameter
$K$.

\section{Conclusion}

The above treatment can be summarized in the three main statements:

1. There is a simple connection between partition functions of gravity
with an extra noncovariant term added to Einstein-Hilbert action
and $SO(4)$ gauge theory (\ref{Fine}), (\ref{Fin}).

2. The noncovariant part of the gravity action in (\ref{Fine}), (\ref{Fin})
is natural to interpret as the gauge-fixing term for a particular gauge
(\ref{gauge}) imposed on the metric tensor.

3. The partition function of $SO(4)$ gauge theory can be brought to the form,
in which the action does not contain fields derivatives, and
the functional integral reduces to the product of independent finite
dimensional integrals at each space points (\ref{Z[A]_1}), (\ref{Z[A]_2}).

The basic method to find gauge/gravity connection relies on the equation
(\ref{Aomega}). By shifting $A_\mu \to A_\mu - \omega_\mu$ it removes
the derivatives either from gauge field or tetrad.
The functional integral without derivatives looks like averaging over
ensemble of uncorrelated random variables. According to the "large numbers
law" the result is weakly sensitive to the distribution of single variables,
being completely determined with a few parameters like mean value and dispersion
accumulating the details. That is why any particular functional measure chosen
for quantum gravity leads to the same standard action for the gauge field
changing the coupling constant(s) only. On the other hand recasting derivatives
onto tetrad yields Einstein-Hilbert gravity action (with fixed gauge)
while the gauge field turns into uncorrelated ensemble and, having been
integrated out, produces the additional local factor for the gravity
measure (\ref{detK}).

The equation (\ref{Aomegapm}) develops this even further completely
removing derivatives both from the tetrad and the field $A_\mu$ without
giving rise to a "gravity". It makes the gluon partition function
to be entirely uncorrelated like that for the lattice with no interaction
between neighbor space points.

There are two comments in order here. First, the replacement
$A_\mu \to A_\mu - \omega_\mu$ removes derivatives only in
the functional integral for the gluon partition function but does
not work for more complex objects such as, say, the Green functions.
The correlations do not disappear for the action with external source.

Second, the divergency of the continuous partition function (\ref{Z[A]_1})
calls for ultraviolet regularization provided with the finite spacing $\Delta x$
and local factor in the measure (\ref{Zeps}). The regularized result (\ref{Z[A]_2}),
that includes apart from these two parameters the bare charge $g$,
could be significantly influenced by subsequent renormalization.
However even if possible corrections are considerable they originate from
short distances and therefore should be in the perturbative region
of the gauge theory.

\section{Appendix}
Here we briefly comment the computation of the integrals
of the form
\begin{equation}
\label{Im}
I_m\,=\,\int de_\mu^A\,\det(e)^m\,e^{-(e_\mu^Ae_\mu^A)^2}
\end{equation}
encountered in the above treatment. It is convenient to start
from a bit more general integral
$$
I\,=\,\int de_i^A\,F(e_i^Ae_k^A),
$$
when the variable $e_i^A$ carries space and color
indices in the intervals $i=1,\ldots, D$, $A=1,\ldots, N_A$
respectively, and $F$ is arbitrary function.
Inserting auxiliary integral over symmetric matrix $g_{ik}$,
$$
I\,=\,\int de_i^A\,F(e_i^Ae_k^A)\,=\,\int \prod_{i\le k}dg_{ik}
\int de_i^A\,\delta(e_i^A e_k^A-g_{ik})\,F(g_{ik}),
$$
where
$$
\delta(X)\,\equiv \,\prod_{i\le k}\delta(X_{ik})
$$
for any symmetric matrix $X$, and using the relation
$$
\delta(C\cdot X\cdot C)=\frac 1{\det C^{D+1}}\,\delta(X)
$$
valid for any symmetric matrix $C$, we transform integral to the form
$$
I\,=\,\int \prod_{i\le k}dg_{ik}\,F(g_{ik})\,
\frac 1{\det g^{\frac{D+1}{2}}}\int de_i^A\,
\delta\bigl(g^{-\frac 12}\,e^Ae^A\,g^{-\frac 12}\,-\,1\bigr),
$$
in which $(e^Ae^A)_{ik}\equiv e^A_ie^A_k$ and positivity of
$g$ assures $g^{\frac 12}$ existence. Changing the variables
$e_i^A=g_{ik}^{\frac 12}\overline e_k^A$,
we finally obtain
\begin{equation}
\label{IFin}
I\,=\,J_{N_A}\,\int \prod_{i\le k}dg_{ik}\,F(g_{ik})\,
\det g^{\frac{N_A-D-1}{2}}
\end{equation}
with the factor
\begin{equation}
\label{JN}
J_{N_A}\,=\,\int d\overline e_i^A\,
\delta(\overline e^A\overline e^A-1)
\end{equation}
independent on $g$ and $F$. The formula (\ref{IFin}) for $N_A=D$
reproduces Jacobian $\det g^{-\frac 12}$ used in the derivation
of the measure (\ref{DeK}).

To find $J_N$ we note firstly that
$$
J_{N+1}\,=\,\int de_k^{N+1}\,\int de_i^A\,\delta\bigl(e_i^A e_k^A -
g_{ik}\bigr)
$$
with $g_{ik}=\delta_{ik}-e_i^{N+1}e_k^{N+1}$, and secondly
that $|\,e_i^{N+1}|\le 1$ in this integral. Since
$\det\,g=1-e_i^{N+1}e_i^{N+1}$ we have the recursion equation
\begin{eqnarray}
J_{N+1}\,&=&\,J_N\,\int de_k^{N+1}\,
\bigl(1-e_i^{N+1}e_i^{N+1}\bigr)^{\frac{N-D-1}{2}}
\nonumber \\
&=&\,J_N\,\Omega_{D}\int_0^1 dr\,r^{D-1}
\bigl(1-r^2\bigr)^{\frac{N-D-1}{2}}
\,=\,J_N\,\Omega_{D}\,B\biggl(\frac 12,\frac{N-D+1}{2}\biggr),
\nonumber
\end{eqnarray}
whose solution reads
$$
J_{D+N}\,=\,\left[\frac 12 \,\Omega_{D}\right]^N
\frac{\Gamma^{N+1}\left(\frac 12\right)}{\Gamma\left(\frac 12+\frac N2\right)}
\,J_D.
$$

Turning back to the integral (\ref{Im}) we see that it can
be calculated through the following chain of equalities
resembling a kind of replica method
\begin{eqnarray}
I_m\,&=&\,J_D\int \prod_{i\le k}dg_{ik}\,e^{-g_{ii}^2}\,
\det g^{\frac{m-1}{2}}\,
=\,\frac{J_D}{J_{D+m}}\int \prod_{A=1}^{D+m} de_i^A\,
e^{-(e_i^Ae_i^A)^2}\,\nonumber \\
&=&\,\frac{J_D}{J_{D+m}}\Omega_{D+m}
\int_0^\infty dr\,r^{D+m-1}e^{-r^4}\,
=\,\frac 14 \frac{J_D}{J_{D+m}}\Omega_{D+m}\,
\Gamma\left(\frac{D+m}{4}\right),
\nonumber
\end{eqnarray}

More general integrals of the form
$\int de_\mu^A\,(Tr g^k)^n\,\det(e)^m\,e^{-(e_\mu^Ae_\mu^A)^2}$
are calculated with the same trick and the help of relations
of the type (\ref{ee}),(\ref{eeee}) with $D\to D+m$,
the factors $\frac{J_4}{J_{N_{2m}}}\Omega_{D+m}$ canceling
in the ratio $\langle (Tr g^k)^n\rangle/\langle 1 \rangle$.

Taking $D=4$ and $m=K$ we get
$$
\langle\,1\,\rangle_K\,=\,\frac{J_4}{J_{K+4}}\Omega_{4(4+K)}
\frac 14\Gamma(K+4),
$$
$$
\langle\,e^4\,\rangle_K\,=\,\frac{J_4}{J_{K+4}}\Omega_{4(4+K)}
\frac 14\Gamma(K+5),
$$
$e^4\equiv (e_i^Ae_i^A)^2$, whereas
$$
g_1^2-g_2\,=\,\frac 32\frac{K+3}{2K+9}\langle\,e^4\,\rangle_K.
$$


\begin{thebibliography}{99}

\bibitem{Plebanski}
Plebanski,
{\em On the separation of Einsteinian substructures,
J.Math.Phys.} {\bf 18} (1977) 2511-2520.

\bibitem{Halpern}
M.B.~Halpern,
{\em Field-strength formulation of quantum chromodynamics},
Phys.\ Rev.\ D{\bf 16} (1977) 1798-1801.

\bibitem{Krasnov}
K.~Krasnov,
{\em Plebanski Formulation of General Relativity: A Practical Introduction},
Gen.\ Rel.\ Grav.\  {\bf 43} (2011) 1
[arXiv:0904.0423 [gr-qc]].

\bibitem{Capovilla}
R.~Capovilla, T.~Jacobson, J.~Dell,
{\em General Relativity without the Metric},
Phys.\ Rev.\ Lett.\ {\bf 63} (1989) 2325-2328.

\bibitem{Lunev_1}
F.~A.~Lunev,
{\em Three dimensional Yang-Mills theory in gauge invariant variables},
Phys.\ Lett.\ B{\bf 295} (1992) 99-103

\bibitem{Lunev_2}
F.~A.~Lunev,
{\em Reformulation of QCD in the language of general relativity},
J.\ Math.\ Phys.\  {\bf 37} (1996) 5351
[arXiv:hep-th/9503133].

\bibitem{Ganor:1995em}
O.~Ganor and J.~Sonnenschein,
{\em The 'dual' variables of Yang-Mills theory and local gauge invariant
variables},
Int.\ J.\ Mod.\ Phys.\  A {\bf 11} (1996) 5701
[arXiv:hep-th/9507036].

\bibitem{Diakonov:2001xg}
D.~Diakonov and V.~Petrov,
{\em Yang-Mills theory as a quantum gravity with 'aether'},
Grav.\ Cosmol.\  {\bf 8} (2002) 33
[arXiv:hep-th/0108097].

\bibitem{BF}
D.~Birmingham, M.~Blau, M.~Rakowski and G.~Thompson,
{\em Topological field theory},
Phys.\ Rep.\ {\bf 209} (1991) 129

\bibitem{Peldan}
P.~Peldan,
{\em Actions for gravity, with generalizations: A Review},
Class.\ Quant.\ Grav.\  {\bf 11} (1994) 1087
[arXiv:gr-qc/9305011].

\bibitem{KP}
N.P.~Konopleva, V.N.~Popov,
{\em Gauge Fields},
Harwood Academic Publishers (1981).

\bibitem{FradVilk}
E. S. Fradkin, G. A. Vilkovisky,
{\em S matrix for gravitational field. ii. local measure, general
relations, elements of renormalization theory},
Phys.\ Rev.\ D{\bf 8} (1973) 4241-4285.

\bibitem{Engle}
J.~Engle, M.~Han and T.~Thiemann,
{\em Canonical path integral measures for Holst and Plebanski gravity. I.
Reduced Phase Space Derivation},
Class.\ Quant.\ Grav.\  {\bf 27} (2010) 245014
[arXiv:0911.3433 [gr-qc]].

\bibitem{Leutwyler}
H.~Leutwyler,
{\em Gravitational Field: Equivalence of Feynman quantization
and canonical quantization},
Phys.\ Rev.\ {\bf 134} (1964) B1155

\bibitem{Lb}
G.~Leibbrandt,
{\em Introduction to noncovariant gauges},
Rev.\ Mod.\ Phys. {\bf 59} (1987) 1067.

\bibitem{Holst}
S.~Holst,
{\em Barbero's Hamiltonian derived from a generalized
Hilbert-Palatini action},
Phys.\ Rev.\ D {\bf 53} (1996) 5966.

\bibitem{Immirzi}
G.~Immirzi,
{\em Real and complex connections for canonical gravity},
Class.\ Quant.\ Grav.\ {\bf 14} (1997) L177.

\bibitem{Rovelli}
C.~Rovelli and T.~Thiemann,
{\em Immirzi parameter in quantum general relativity},
Phys.\ Rev.\ D {\bf 57} (1998) 1009.

\end{thebibliography}
\end{document}